\documentstyle[aps,psfig,prl,floats]{revtex}
\begin{document}
\topmargin=0.0cm
\firstfigfalse\firsttabfalse

\twocolumn[\hsize\textwidth\columnwidth\hsize\csname@twocolumnfalse\endcsname

\title{Effects of disorder on the optical properties of the\\
(Zn,Mg)(S,Se) quaternary alloy}

\author{Antonino Marco Saitta,$^{1,}$\cite{present} 
Stefano de Gironcoli,$^{1,}$\cite{sdg} and Stefano Baroni$^{1,2,}$\cite{sb} }

\address{$^1$INFM -- Istituto Nazionale per la Fisica della Materia
and \\ SISSA -- Scuola Internazionale Superiore di Studi Avanzati,
Via Beirut 2-4, I-34014 Trieste, Italy \\ $^2$CECAM -- Centre
Europ\'een de Calcul Atomique et Mol\'eculaire, ENSL, 46~All\'ee
d'Italie, 69364 Lyon Cedex 07, France} 

\date{\today}
\maketitle

\begin{abstract} The electronic and optical properties of
(Zn,Mg)(S,Se) wide-gap solid solutions are studied using {\it ab
initio} techniques and starting from the previously determined
atomistic structure of the alloy. Compositional disorder is shown to
close substantially the gap with respect to the predictions of the
virtual-crystal approximation. The bowing of the fundamental gap
vs. composition predicted by our calculations is in very good
agreement with experiments available for the Zn(S,Se) pseudo-binary
alloy. At temperatures typical for MBE growth, the quaternary alloy
displays a rather large amount of short-range order whose effect is to
slightly but unmistakably open the gap. Our results agree well with
recent experimental data for the quaternary alloy. \end{abstract}

\pacs{PACS numbers: 
61.43.-j 
61.66.Dk 
71.23.-k 
71.55.Gs} 

]
\narrowtext

In the last decade much effort has been devoted to the study of
semiconducting materials capable of operating in the short wavelength
optical range. The main goal of these efforts is the realization of
light-emitting and laser diodes encompassing the entire visible-light
window and, in perspective, the industrial-scale production of
high-density storage optical disks and light sources for full-color
displays. II-VI and nitride semiconductors are the most promising
among these materials. Among II-VI semiconductors, ZnSe-based
materials have a special importance \cite{II_VI_Review}, and
quaternary ${\rm Zn}_x{\rm Mg}_{1-x}{\rm S}_y{\rm Se}_{1-y}$ alloys
are commonly used as cladding layers in II-VI blue-green laser
diodes. This system---which was introduced a few years
ago~\cite{ZnMgSSe}---has the desirable property that its lattice
parameter, $a_0$, and its fundamental band gap, $E_g$, can be tuned
fairly independently by acting on the concentrations $x$ and $y$. In
spite of the technological importance of this material, extensive
experimental studies of its electronic and optical properties over a
wide range of compositions and of their dependence on the structural
and thermodynamic equilibrium properties are still lacking.

We recently reported~\cite{noi} on a theoretical study of the
thermodynamical properties of this system, performed using
state-of-the-art {\it ab-initio} methods. In that study the phase
diagram of the quaternary alloy was determined, and the homogeneous
alloy was found to be stable against segregation or the formation of
ordered structures at temperatures typical of molecular-beam epitaxy
(MBE) growth ($\approx$ 550 K). Nevertheless, a large amount of
short-range order (SRO) characterized by the occurrence of Zn-Se and
Mg-S clustering among first-nearest neighbors was found to occur and
to persist even at very high temperatures ($\approx$ 1700 K).

In this Letter we report on the first extensive theoretical study of
the electronic and optical properties of (Zn,Mg)(S,Se) alloys both in
the case of the experimentally well studied pseudo-binary ${\rm
ZnS}_x{\rm Se}_{1-x}$ system, and for the quaternary solid solution in
conditions of lattice-matching to GaAs substrates. In the latter case
the presence of the short range correlations will be shown to induce a
small but significant opening of the fundamental band gap.

Our work is based on density-functional theory (DFT) within the local
density approximation (LDA), and the plane-wave pseudopotential
method. All the technical details of our calculations are the same as
in Ref. \cite{noi}. DFT-LDA is known to underestimate electronic
excitation energies, and many-body corrections should be considered,
{\it e.g.} using the GW method \cite{GW}. In spite of this, the
agreement between the {\it shape} of the band-structure predicted by
the LDA and that measured experimentally is usually satisfactory, and
semi-quantitative agreement can be simply obtained by a rigid upward
shift of the conduction bands, using the so-called {\it scissor
operator}.

Ideally, the study of the electronic properties of an alloy should
proceed in three steps: {\it i)} the determination of the equilibrium
structural properties of the system, including the relevant atomic
correlations, as we obtained and discussed in Ref.~\cite{noi}; {\it
ii)} the calculation of the alloy band structure, taking into account
the atomistic structure previously determined; {\it iii)} the
inclusion of electronic correlations. It is still very hard to combine
many-body calculations \cite{GW} with a proper description of
disorder, and in the following we assume that disorder and
quasi-particle effects can be treated independently: we deal
explicitly with compositional disorder at the LDA level, and we add
semi-empirically the many-body corrections by a simple interpolation
of the scissor operators appropriate for the pure compounds.

Starting from the structural data determined in Ref. \cite{noi}, we
have calculated the electronic properties of the alloys using the {\it
special quasi-random structures} approach~\cite{SQS}. The latter
is based on the observation that---for any given composition---atomic
disorder mainly affects the electronic properties of an alloy through
the short-range atomic structure. A disordered solid solution can thus
be mimicked by using reasonably small supercells that reproduce the
alloy SRO in the first few (typically, four or five) shells of
neighbors. We have generalized the original method so as to account
for any given short-range order~\cite{Mader}, arbitrary compositions
and double sub-lattice disorder.  We have verified that atomic
correlations beyond the fourth shell of neighbors affect only
negligibly the band structure of the alloys presently studied.  The
supercell structures used in our simulations were thus obtained by a
simulated-annealing procedure by which we searched among 64-atom
simple-cubic structures the ones which give the most similar pair
correlations up to the fourth shell of neighbors, as compared to those
obtained from Monte Carlo simulations \cite{noi}.

\begin{figure} \centerline{\psfig{figure=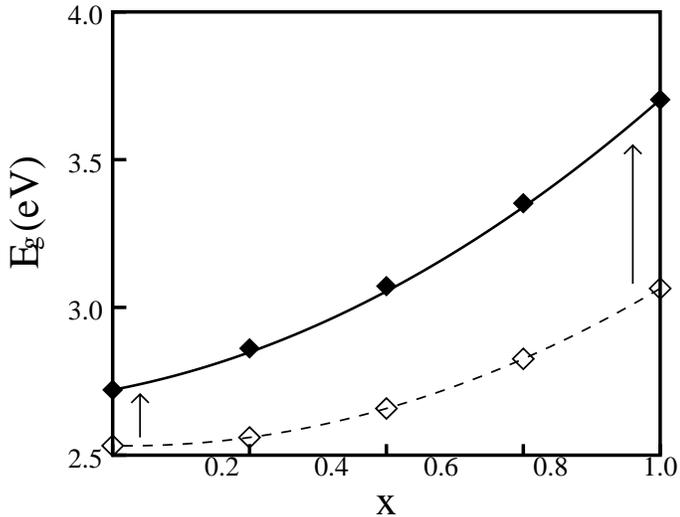,width=10cm}} \small
\def\baselinestretch{1.2} \caption{Energy gap of the ${\rm Zn(S,Se)}$
pseudo-binary alloy as a function of sulphur concentration, $x$. Solid
line: experimental data from Ref.[7]. Open diamonds: our theoretical
results obtained neglecting many-body effects. Dashed line: quadratic
interpolation to our theoretical results. Filled diamonds: theoretical
results semi-empirically corrected for quasi-particle effects (see
text).} \label{gapZSS} \end{figure}


The electronic properties of the ${\rm ZnS}_x{\rm Se}_{1-x}$ pseudo-binary
alloy, which are well characterized experimentally, have been studied
by the above techniques, at three different concentrations, {\it i.e.}
$x=$0.25, 0.50, 0.75. In Fig.~\ref{gapZSS} we display our results for
the fundamental band gap of the alloy and compare them with available
experimental data \cite{ebina}. As previously discussed, LDA calculations
systematically underestimate the optical gap. Adding a simple linear
interpolation of the errors done at the pure-compound extremes, brings
our theoretical predictions in very good agreement with experiments.
Upon this semi-empirical correction, our data can be accurately
described by the formula: \begin{equation} E_g(x) = E_g^{\rm
ZnSe}(1-x)+E_g^{\rm ZnS}x+bx(1-x), \end{equation} where the curvature,
commonly known as {\em bowing parameter}, is $b\approx 0.56~ \rm
eV$. Our result is in good agreement with experiments \cite{ebina,bowing},
according to which $0.41 {\rm ~eV} \leq b \leq 0.68 \rm ~eV$. It is
worth noting that calculations done in the {\it virtual crystal
approximation} ({\it i.e.} by neglecting disorder and averaging the
atomic pseudopotentials in the anionic sub-lattice) would predict a
small and {\em negative} bowing parameter: $b\approx -0.1 \rm
~eV$. The accuracy by which our calculations predict the non-linear
dependence of the band gap upon composition in the pseudo-binary ${\rm
ZnS}_x{\rm Se}_{1-x}$ alloy indicates an almost linear dependence of
quasi-particle effects on the concentration, which is encouraging for
the study of the electronic properties of the quaternary alloy, where
the available experimental data are scarcer.

\begin{figure}
\centerline{\psfig{figure=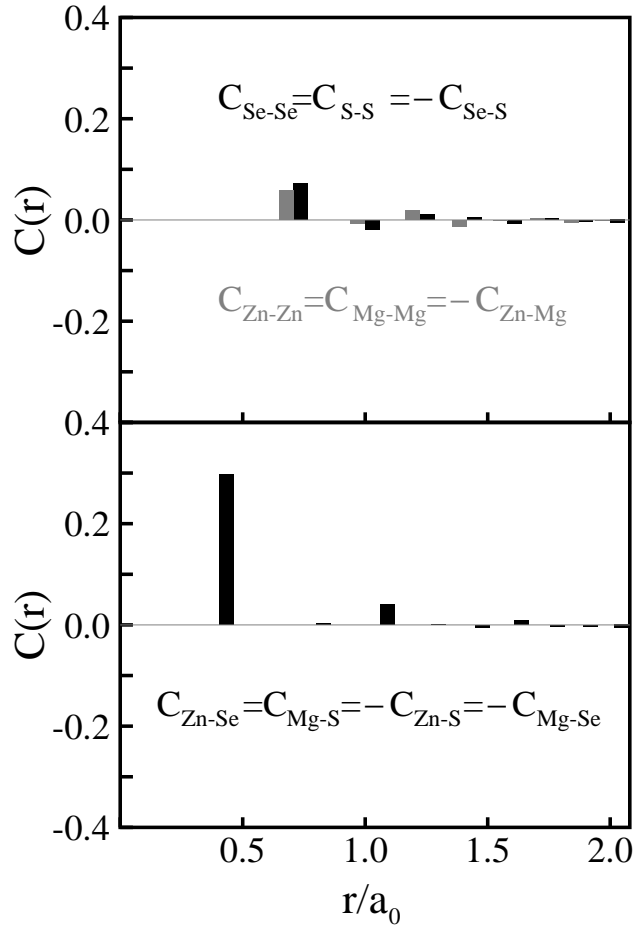,width=10.0cm}} \small
\def\baselinestretch{1.2} \caption{Interatomic correlation functions
in $\rm Zn_{1\over 2} Mg_{1\over 2} S_{1\over 2} Se_{1\over 2}$ solid
solutions (see Eq. (2)), as calculated in Ref. [3] for T= 550 K.}
\end{figure}

The quaternary alloys displays a large amount of short-range
order. This is illustrated in Fig. 2 where we display the atomic
correlation functions calculated in Ref. \cite{noi} at a temperature
of 550 K and for $x=y={1\over 2}$. The correlation function between
two atomic species, X and Y, is defined as: \begin{equation} C_{\rm
XY}({\bf r}) = \langle \xi({\bf r})\eta(0) \rangle -\langle \xi
\rangle\langle\eta\rangle, \end{equation} where $\xi({\bf r})$ and
$\eta({\bf r})$ are integer random variables whose value is 1 if the
atomic site at position $\bf r$ is occupied by an atom of species X or
Y respectively, and 0 otherwise; $\langle\cdot\rangle$ indicates the
average over disorder; and $\langle \xi \rangle = x$ and
$\langle \eta \rangle = y $ are the concentrations of the X
and Y atomic species. The large positive value of the Zn--Se and Mg--S
first-nearest neighbor correlations indicates a strong tendency to
form ZnSe and MgS local clusters. This kind of clustering acts so as
to reduce the elastic energy of the system because ZnSe and MgS have
lattice parameters similar to each other and to that of the alloy
($a_{\rm ZnSe} = 10.57 a.u.$ and $a_{\rm MgS} = 10.79 a.u.$), 
but very different from those of ZnS and MgSe ($a_{\rm ZnS} = 10.09 a.u.$ 
and $a_{\rm MgSe} = 11.32 a.u.$). This
tendency also shows in the second-nearest neighbor correlations which
favor like cations and anions.

The electronic structure of the quaternary alloy has been studied for
three different concentration pairs along the line of lattice-matching
to ${\rm GaAs}$ ({\it i.e.} for pairs of concentrations, $x$ and $y$,
such that the average lattice parameter of the alloy is the same as
that of GaAs). 

\begin{table}\small \def\baselinestretch{1.2} \caption{Energy gap {\it
vs.} compositions in (Zn,Mg)(S,Se). `Virtual': theoretical results
obtained within the virtual-crystal approximation. `Random':
super-cell calculations for the random alloy. `SRO': supercell
calculations done including SRO, but neglecting many-body
effects. `Corrected SRO': as above, but corrected for quasi-particle
effects (see text). `Expt': experimental data from Ref. [9].}
\smallskip
\begin{tabular}{c c c c c c c} $x$ & $y$ & Virtual & Random & SRO &
Corrected SRO & Expt \\ \hline 0.50 & 0.50 & 2.96 & 2.57 & 2.64 & 3.52
& 3.56 \\ 0.75 & 0.13 & 2.68 & 2.48 & 2.50 & 3.00 & 3.00 \\ 0.84 &
0.00 & 2.55 & 2.43 & 2.44 & 2.78 & 2.82 \\ \end{tabular}
\label{t:gaptot} \end{table}

In Table \ref{t:gaptot} we report the values of the fundamental gap as
calculated taking into account SRO, neglecting SRO, and in the
virtual-crystal approximation, and we compare these results with
recent experimental data \cite{Egap_exp}. Quasi-particle effects are
taken into account by the same kind of semi-empirical corrections
described in the case of the pseudo-binary alloy. In the present
quaternary case, the values of scissor operators for the four pure
compounds are bi-linearly interpolated in-between. Our calculated
values for the energy gaps are in very good agreement with experiments
(within 0.04 eV).  As it was the case for the Zn(S,Se) pseudo-binary
alloy, the predictions of the virtual-crystal approximation are rather
poor, resulting in too large a gap. The effect of SRO, as compared to
the perfectly random solution, is to slightly open back the gap. The
effect is small, but sizeable when disorder is maximum ({\it i.e.} for
$x=y={1\over 2}$). The effects of SRO on the alloy band structure can
be understood qualitatively considering the actual alloy as a
perturbation with respect to the appropriate virtual crystal (VC), as done
by Baldereschi and Maschke \cite{BaldereschiMaschke75} in an early
study of the Ga$_x$In$_{1-x}$P alloy. Consider the actual alloy
potential as a perturbation with respect to the VC potential:
\begin{equation} V_{alloy}({\bf r}) = V_{VC} ({\bf r}) + \Delta V ({\bf r}).
\end{equation} To second order in the perturbation, the alloy band
structure is given by \begin{equation} \epsilon_n({\bf k}) =
\epsilon_n^{0} ({\bf k}) + \sum_{n',{\bf k'}} \frac{|\langle n, {\bf
k}| \Delta V | n', {\bf k'}\rangle|^2} { \epsilon_n^{0} ({\bf k}) -
\epsilon_{n'}^{0} ({\bf k'})}, \end{equation} where $| n, {\bf k}
\rangle $ are the virtual-crystal eigenvectors with energy
$\epsilon^{0}_n({\bf k})$, for the appropriate given composition pair
$(x,y)$. Since occupied (empty) states mostly interact with the other
occupied (empty) states, that are closer in energy, the effect of the
perturbation is to push the band edges in the gap region, thus
reducing the virtual crystal gap, in agreement with our findings for
${\rm Zn_xMg_{1-x}S_ySe_{1-y}}$, random and correlated, quaternary alloy
as well as for ${\rm ZnS_xSe_{1-x}}$ alloy. In the presence of SRO,
this result can be generalized by showing that the magnitude of the
gap reduction depends quadratically on the strength of the localized
perturbations which transform the virtual crystal into the actual
alloy, multiplied by the appropriate atomic correlation functions.
From Fig.\ 2 it can be seen that the main difference between the
correlated and the random alloy is that the former has a larger number
of bond-length preserving Zn--Se and Mg--S nearest-neighbor pairs and a
reduced number of bond-stretching Zn--S and Mg--Se nearest-neighbor
pairs relative to the random alloy. The former type of perturbation of
the virtual crystal is probably weaker than the latter, thus
explaining the smaller closure of the VC gap in the correlated alloy
as compared to the random case.

In order to better characterize the optical gap of the alloy, we
consider the spectral weight: \begin{equation} A(E,{\bf k}) = \sum_n
\bigl | \langle \psi_n | P({\bf k}) | \psi_n \rangle \bigr |^2
\delta(E-\epsilon_n), \end{equation} where $\psi_n$ and $\epsilon_n$
are the wave-functions and energy levels of the supercell, and $P({\bf
k})$ is the projector over the states of quasi-momentum $\bf k$. We
have found that the alloy wave-functions at the top of the valence and
at the bottom of the conduction bands actually have a strong $\Gamma$
character, thus indicating that the direct-gap nature of the pure
materials is thus conserved also in the alloy. An analysis of the
different atomic contributions to the DOS has been carried out by
projecting the super-cell eigenfunctions onto the atomic-like
localized orbitals. We found that the states at the top of the valence
band have a strong anionic character (26\% centered on S and 67\% on
Se). The anionic character of the top of the valence band is a direct
consequence of the ionic character of the alloy, while the larger
contribution of the Se atomic orbitals is consistent with Harrison's
tight binding tables~\cite{harrison} which indicate that the top of
the valence band states of ${\rm S}$-compounds lies about 0.5 eV below
the analogous states of the corresponding ${\rm Se}$-compounds. The
low-lying conduction states are rather delocalized, with contributions
from all the four atomic species and a slight prevalence of cationic
character (28\% Zn and 33\% Mg).

A more complete account of this work can be found in the Ph.D. thesis
of one of us~\cite{PhD}. We thank A. Franciosi and S. Rubini for
prompting our interest in this problem and for many fruitful
discussions. This work has been partly done within the {\it Iniziativa
Trasversale Calcolo Parallelo} of the INFM.


\end{document}